\documentstyle[12pt]{article}
\def\deg{\ifmmode{^{\circ}}\else ${^{\circ}}$\fi}

\newcommand{\bi}{\begin{itemize}}
\newcommand{\ei}{\end{itemize}}
\def\ed{\end{document}}
\begin{document}
\newcommand{\lqcd  }{ \mbox{$\Lambda_{QCD} $} }
\newcommand{\lqcdsq  }{ \mbox{$\Lambda^2_{QCD} $} }
\begin{titlepage}
\begin{flushright}
{\sl NUB-3074/93-Th}\\
hep-ph/9310217\\
{\sl September 1993}
\end{flushright}
\vskip 0.5in
\begin{center}
{\Large\bf  Can $\gamma \gamma \rightarrow Z_L Z_L$ serve as a probe of the
electroweak symmetry breaking sector ? }\\[.5in]
{\sc Rogerio Rosenfeld}\footnote{e-mail: rosenfeld@neuhep.hex.neu.edu}
\\[.5in]
{\sl Department of Physics}\\
{\sl Northeastern University}\\
{\sl Boston, MA 02115}
\end{center}
\vskip 1in
\begin{abstract}
We investigate the feasibility of distinguishing among different models of
electroweak symmetry breaking by studying the process
$\gamma \gamma \rightarrow Z_L Z_L$ at photon colliders. For models with a low
mass Higgs-like scalar resonance, the $s$-channel contribution provides
a distinct resonance structure and large cross sections.
However, the absence of a resonance structure in the cross section for
the above process does not discriminate in principle between
the existence of either a heavy Higgs boson or a vector resonance of a
non-linearly realized symmetry.
\end{abstract}
\end{titlepage}
\clearpage
\setcounter{page}{2}
\section{Introduction}
\hspace*{\parindent}

The Standard Model has been tested to a remarkable accuracy at LEP
but there is still very little to be said about its electroweak
symmetry breaking (EWSB) sector apart from a lower limit on the Higgs
boson mass, $M_H > 63.5$ GeV \cite{LEP} .
On the theoretical side, the existence of a Higgs boson
by itself makes it
difficult to understand the fine-tuning that protects its mass to be
as large as the Planck mass.
There are two possible ways to avoid this so-called hierarchy
problem \cite{hierarchy}. One could either have a supersymmetric theory,
in which quantum
corrections to the Higgs mass are only logarithmic instead of quadratic,
or one could have
a strongly interacting underlying theory which has the EWSB sector
as its low-energy effective theory.
In spite of some new indirect evidence in favor of supersymmetric
models \cite{Amaldi}, here we will concentrate on the second approach,
which has rich
experimental consequences at the TeV scale that could be tested at the
SSC and LHC.

The presence of this strongly interacting sector would manifest itself
primarily in the scattering of the longitudinal components of the weak
gauge bosons $W_L,Z_L$, which at high energies behave as the pseudo
Nambu-Goldstone bosons originating from the global
symmetry breaking occurring in the EWSB sector,
$SU(2)_L \times SU(2)_R  \rightarrow SU(2)_V $.
The residual $SU(2)_V$ symmetry is responsible for keeping $\rho =
M_W/(M_Z \cos \theta) =  1$ at tree level.
Therefore, $W_L$'s and $Z_L$'s behave as techni-pions of this underlying
strong dynamics and we can describe their interactions by using chiral
lagrangian techniques. If this scenario is correct, it is very plausible
that resonances would also play a role in describing physical processes.
In fact, amplitudes derived solely from the pion sector of a chiral lagrangian
grow with energy and violate unitarity rather quickly; the presence of
resonances tend to unitarize these amplitudes.
At this point, there are two types of resonances that could appear : scalar
and vector resonances. The usual Higgs sector is an example of the former
choice; however, in QCD we know that vector resonances are more important in
restoring unitarity, {\it i.e.}, in saturating scattering amplitudes.
Therefore,
we should keep an open mind with respect to what type of resonance would
show up in $W_L W_L$ scattering.
Both types of resonances can be nicely described in terms of chiral
lagrangians.
A detailed study of consequences of chiral lagrangians with either
scalar or vector resonances (and of other models as well) in
$W_L W_L$ scattering was performed recently by J. Bagger {\it et al.}
\cite{Bagger} .

In this paper we explore the possibility of distinguishing between
chiral lagrangians with scalar or vector resonances by studying their
contribution to the process
$ \gamma \gamma \rightarrow
Z_L Z_L $.
The suggestion of obtaining high energy $\gamma$ beams from backscattered laser
and the smallness of the above process in the Standard Model makes it
in principle a good window to explore physics beyond the Standard Model.
There has been some recent activity in the study of $ \gamma \gamma \rightarrow
Z_L Z_L $.
Abbasabadi et al. \cite{Abba} used a K-matrix
unitarized amplitude to study this process in the Standard Model, i.e.,
a model with a scalar resonance.
They found a substantial correction to the same calculation without
unitarization \cite{Boos} only for $M_H > 5$ TeV, in accord to the fact that
the Standard Model amplitudes violate unitarity for $M_H > {\cal O}
(1 \mbox{TeV} ) $.
This process was also considered in the context of chiral lagrangians without
any resonances in ref. \cite{Herrero} and in the context of anomalous gauge
couplings in ref. \cite{Belanger}.

In the next section we compute the amplitudes for $\gamma \gamma \rightarrow
Z_L Z_L $ in chiral lagrangian models with no resonance, a scalar resonance
and a vector resonance. In section 3 we compare the cross sections arising
from these different models. In section 4 we examine the possibility of
distinguishing these models in a realistic $\gamma \gamma$ machine and
we conclude in section 5.

\section{Models and amplitudes}

\subsection{ Chiral lagrangian without resonances}
\hspace*{\parindent}

This model describes the interaction of the
pseudo Nambu-Goldstone bosons $w^{a} (a=1,2,3)$ among themselves
and  the photon field $A_\mu$ given by the lagrangian
\begin{equation}
{\cal L} = \frac{v^2}{4} Tr[ D_\mu U D^\mu U^\dagger ]  - \frac{1}{4}
( F_{\mu \nu} )^2 ,
\label{eq:chiral}
\end{equation}
where
\begin{equation}
U = \exp [2 i w/v] \;\;\; , \;\;\; w = w^a \tau^a/2 \;\;\;,\;\;\;
v = 246 \; \mbox{GeV} ,
\end{equation}
and the covariant derivative is given by :
\begin{equation}
D_\mu U = \partial_\mu  U + i e A_\mu [Q,U]  .
\label{eq:covariant}
\end{equation}

A few words must be said about the choice for the charge matrix $Q$.
The chiral lagrangian possesses the symmetries of the underlying theory
involving fermions. In the general case where these symmetries are
local $G_L \times G_R$ and the fundamental fermions transform as
\begin{equation}
\psi_L \rightarrow L \psi_L  \;\;\; , \;\;\; \psi_R \rightarrow R \psi_R
\end{equation}
where $L \in G_L$ and $R \in G_R$, the matrix $U$ transforms as :
\begin{equation}
U \rightarrow L U R^\dagger  .
\end{equation}
Introducing the gauge fields $l_\mu$ and $r_\mu$ transforming as
\begin{equation}
l_\mu \rightarrow L l_\mu L^{\dagger} + i (\partial_\mu L) L{^\dagger}  \;\;\;
,
\;\;\;
r_\mu \rightarrow R r_\mu R^{\dagger} + i (\partial_\mu R) R^{\dagger}
\end{equation}
the covariant derivative becomes :
\begin{equation}
D_\mu U = \partial_\mu  U + i l_\mu U - i U r_\mu  .
\end{equation}
In the case of QCD with only up and down quarks, the electromagnetic
interactions are introduced by the identifications :
\begin{equation}
l_\mu = r_\mu = e A_\mu Q   \;\;\;, \;\;\;
Q = \left( \begin{array}{cc}
            2/3  &  0  \\   0  &  -1/3
            \end{array}
     \right)  ,
\end{equation}
recovering Eq. (\ref{eq:covariant}).

In the case of the Standard Model, we introduce $SU(2)_L \times U(1)_Y $
gauge interactions by choosing :
\begin{eqnarray}
l_\mu &=& g' \frac{Y_L}{2} B_\mu + g \frac{\tau^a}{2} W_\mu^a \nonumber \\
r_\mu &=& g' \frac{Y_R}{2} B_\mu
\end{eqnarray}
which correspond to a charge matrix :
\begin{equation}
Q = \left( \begin{array}{cc}
            1/2  &  0  \\   0  &  -1/2
            \end{array}
     \right)   +
\frac{Y_L}{2} \; \left( \begin{array}{cc}
            1  &  0  \\   0  &  1
            \end{array}
     \right)  ,
\label{eq:charge}
\end{equation}

This charge matrix should reflect the electroweak quantum numbers of
the underlying techni-fermions; it should be really considered as a
free parameter
of the model. Here we'll parametrize the charge matrix by
$Y_L$, the hypercharge of the techni-fermion doublet.
We have checked that all interactions involving only photons and $w$ fields
described by the lagrangian Eq.(\ref{eq:chiral} ) are independent of $Y_L$.
For the general case of $Y_L \neq 0$, extra heavy fermions would be
required to cancel undesirable anomalies.

We are interested in the process $\gamma \gamma \rightarrow Z_L Z_L$.
Using the equivalence theorem, which was proved to all orders in the weak
coupling constant in any $R_\xi$ gauge \cite{ET}, we have for the scattering
amplitude at a given center-of-mass energy $\sqrt{s}$ (denoting $z = w^3$) :
\begin{equation}
A( \gamma \gamma \rightarrow Z_L Z_L ) = A( \gamma \gamma \rightarrow z z )
+ {\cal O} (M_Z^2/s).
\end{equation}
In the following we'll write :
\begin{equation}
 A( \gamma \gamma \rightarrow z z ) =  \epsilon_1^\mu \epsilon_2^\nu
\; T_{\mu \nu} ,
\label{eq:amplitude}
\end{equation}
where $\epsilon_{1,2}$ are the polarization $4$-vectors of the initial
photons.

The calculation of the process $\gamma \gamma \rightarrow z z $ is
straightforward but tedious. From the  lagrangian Eq.(\ref{eq:chiral} )
one obtains the relevant couplings $\gamma w w$, $\gamma \gamma w w $,
$w w w w$,
$\gamma \gamma w w w w$ and $\gamma w w w w$ ($ w = w^{\pm},z$) that enter
in the computation
of the relevant one-loop Feynman diagrams (there are no tree-level
contributions in this model). The result for this non-resonance (NR) model
is \cite{Don,Bij} :
\begin{equation}
T_{\mu \nu}^{NR} =
- i \frac{\alpha}{2 \pi}
\left( g_{\mu \nu} - \frac{k_{1 \nu} k_{2 \mu} }{k_1 \cdot k_2} \right)
\frac{s}{ v^2}  ,
\end{equation}
where $k_1 $ and $k_2$ are momentum $4$-vectors associated with the photons.
The result is finite because there is no tree-level contribution to this
process at ${\cal O}(p^4) $ that would absorb the divergencies.
Noticing that in this model,
\begin{equation}
A_{NR}(w^+ w^- \rightarrow z z) = i s/v^2 ,
\label{eq:LET}
\end{equation}
we can write :
\begin{equation}
T_{\mu \nu}^{NR} =
- \frac{\alpha}{2 \pi} \left( g_{\mu \nu} - \frac{k_{1 \nu} k_{2 \mu} }
{k_1 \cdot k_2} \right)
A_{NR}(w^+ w^- \rightarrow z z),
\label{eq:result}
\end{equation}
which can be interpreted as a rescattering $\gamma \gamma \rightarrow
w^+ w^- \rightarrow z z$ .

This simple model should not be trusted at energies
$\sqrt{s} > {\cal O}(4 \pi v)$, where unitarity is violated in some
amplitudes for $ww$ scattering and all the terms in the derivative
expansion become of the same order. The effects of higher order terms can
be estimated by introducing resonances that saturate the scattering amplitudes
or by unitarizing these amplitudes in an {\it ad hoc} manner.
The latter approach was used by the authors of ref. \cite{Don,Dob} in the case
of $ \gamma \gamma \rightarrow \pi^0 \pi^0 $.

The result Eq.(\ref{eq:result}) will serve as an {\it ansatz} to study the
influence of
resonances on the process $ \gamma \gamma \rightarrow Z_L Z_L $, {\it i.e.},
we will assume that this relation is valid in the other models discussed
below as well.
In the next subsections we'll examine two classes of models : a Higgs-like
model
with a scalar resonance and a technicolor-like model with a vector resonance.

\subsection{ Chiral lagrangian with a scalar resonance}
\hspace*{\parindent}

Here we consider the most general model of a scalar particle $H$ coupled in
a chiral invariant way \cite{DonVal} :
\begin{equation}
{\cal L}_S = \frac{v^2}{4} Tr[ D_\mu U D^\mu U^\dagger ]  - \frac{1}{4}
( F_{\mu \nu} )^2  + \frac{1}{2} \partial_\mu H \partial^\mu H -
\frac{1}{2} M_H^2 H^2 + \frac{1}{2} g_H v H  Tr[ D_\mu U D^\mu U^\dagger ] .
\end{equation}

The scalar resonance is parametrized by its mass $M_H$ and by a coupling
constant $g_H$. Its width is given by :
\begin{equation}
\Gamma_H = \frac{3 g_H^2 M_H^3}{32 \pi v^2}
\end{equation}
and the relevant scattering amplitude in this model reads :
\begin{equation}
A_H(w^+ w^- \rightarrow z z ) = i \frac{s}{v^2} \left[ 1 - \frac{g_H^2 s}{s -
M_H^2 + i M_H \Gamma_H} \right] .
\end{equation}
Notice that $H$ reduces to the Standard Model Higgs boson for $g_H = 1$. In
what
follows we'll restrict ourselves to the case $g_H=1$. We don't expect any major
differences for other cases.

{}From Eq.(\ref{eq:result}) we find the contribution of a generic
scalar resonance to $ \gamma \gamma \rightarrow z z$ :
\begin{equation}
T_{\mu \nu}^{H} =
- i \frac{\alpha s}{2 \pi v^2} \left( g_{\mu \nu} - \frac{k_{1 \nu} k_{2 \mu} }
{k_1 \cdot k_2} \right)
 \left[ 1 - \frac{g_H^2 s}{s -M_H^2 + i M_H \Gamma_H} \right]  .
\end{equation}
For $g_H=1$, this agrees with the result of Ref. \cite{Boos}, which
was obtained in the context of the Standard Model but keeping only
couplings of enhanced electroweak strength, i.e., neglecting couplings
of the order ${\cal O} (g^2) $ compared to  ${\cal O} (g^2 M_H^2/M_W^2) $ .

\subsection{ Chiral lagrangian with a vector resonance }
\hspace*{\parindent}

We introduce a vector resonance in the chiral lagrangian as a gauge vector
boson of a local $SU(2)$ group by means of
the so-called hidden symmetry approach \cite{Bando}, which has been
successful in describing the properties of vector mesons in QCD. It was shown
to be equivalent to other different approaches in Ref. \cite{eckeretal}.

We parametrize the matrix $U$ by :
\begin{equation}
U = \xi_L^\dagger \xi_R
\end{equation}
with the following transformations under $ G_L \times G_R \times SU(2) $ :
\begin{eqnarray}
\xi_L &\rightarrow& h \xi_L L^\dagger  \nonumber  \\
\xi_R &\rightarrow& h \xi_R R^\dagger             \\
U &\rightarrow&     L U R^\dagger          \nonumber
\end{eqnarray}
where  $L \in G_L \;,\;R \in G_R$ and $h \in SU(2)$ . Let's assume for the
moment that $G_L$ and $G_R$ are global groups; external gauge fields are
easily introduced by gauging the appropriate subgroups.

The vector resonance $V_\mu = V_\mu^a \frac{\tau^a}{2}$ transforms as a
$SU(2)$ gauge field :
\begin{equation}
V_\mu \rightarrow  h V_\mu h^\dagger + \frac{i}{g''} h \partial_\mu h^\dagger ,
\end{equation}
where $g''$ is the coupling constant associated with the local $SU(2)$.
We define the covariant derivative
\begin{equation}
D_\mu \xi_{L,R} = \partial_\mu \xi_{L,R} - i g'' V_\mu \xi_{L,R}
\end{equation}
such that
\begin{equation}
D_\mu \xi_L \rightarrow  h D_\mu \xi_L  L^\dagger  \;\;\;,\;\;\;
D_\mu \xi_R \rightarrow  h D_\mu \xi_R  R^\dagger
\end{equation}

The building blocks to construct an invariant lagrangian are :
\begin{equation}
\alpha_{L \mu} = (D_\mu \xi_L)  \xi_L^\dagger \;\;\; , \;\;\;
\alpha_{R \mu} = (D_\mu \xi_R)  \xi_R^\dagger
\end{equation}
which transform as
\begin{equation}
\alpha_{L \mu} \rightarrow h \alpha_{L \mu} h^\dagger  \;\;\;,\;\;\;
\alpha_{R \mu} \rightarrow h \alpha_{R \mu} h^\dagger  .
\end{equation}

Finally , the most general
lowest order lagrangian which respects the parity-like symmetry
$L \leftrightarrow R$ is given by :
\begin{equation}
{\cal L}_V = -\frac{1}{4} Tr[ (V_{\mu \nu})^2 ] -
\frac{v^2}{4} Tr[ ( \alpha_{L \mu} - \alpha_{R \mu} )^2 ] - a \frac{v^2}{4}
 Tr[ ( \alpha_{L \mu} + \alpha_{R \mu} )^2 ] ,
\end{equation}
where $V_{\mu \nu}$ is the non-abelian field-strength for the vector resonance.
There are two free parameters in this lagrangian, namely, $g''$ and $a$.
It can be shown that in the limit $g'' \rightarrow \infty$, the kinetic term
for the vector resonance vanishes and the resonance becomes an auxiliary
field which is
eliminated by its equation of motion, which in turn reduces the above
lagrangian
to the usual non-linear $\sigma$ model lagrangian :
\begin{equation}
{\cal L}_V  \stackrel{g'' \rightarrow \infty}{\rightarrow}
\frac{v^2}{4} Tr[ \partial_\mu U \partial^\mu U^\dagger ]   .
\end{equation}

The mass and width of the vector resonance in this model can be derived
from the lagrangian
after using an $SU(2)$ gauge transformation to set $\xi_L^\dagger =
\xi_R = \exp [ i w/v] $ (unitary gauge):
\begin{eqnarray}
M_V^2 = a g''^2 v^2  \nonumber  \\
\Gamma_V = \frac{a M_V^3}{192 \pi v^2}
\end{eqnarray}
and the scattering amplitude of interest to us is given by \cite{vector}:
\begin{equation}
-i A_V(w^+ w^- \rightarrow z z) = \frac{s}{4 v^2} (4 - 3 a ) +
\frac{a M_V^2}{4 v^2} \left[ \frac{u - s}{t - M_V^2} +
\frac{t - s}{u - M_V^2}  \right]   .
\end{equation}
Notice that the above amplitude reduces to Eq.(\ref{eq:LET}) in the limit
$M_V^2 \gg t,u$, reproducing the low-energy theorems.

However, the parity-like operation $L \leftrightarrow R$ is not a symmetry
of the underlying theory. It corresponds to a symmetry of the theory under
$w(\vec{x},t) \rightarrow -w(\vec{x},t)$ , which forbids transitions between
even and odd numbers of the pseudo-Goldstone bosons $w$. However, parity
conservation implies in the symmetry $w(\vec{x},t) \rightarrow -w(-\vec{x},t)$
and it is possible to write down parity-conserving terms in the lagrangian
that violate the  $L \leftrightarrow R$ symmetry \cite{extra}.
In QCD, these terms describe processes like $\rho, \omega \rightarrow
\pi \gamma $. An analogue term in the model we are studying is
uniquely (up to total derivatives) determined to be \cite{Chivukula}  :
\begin{equation}
{\cal L}_{V \gamma w } = \kappa \frac{e g''}{v} \epsilon^{\mu \nu \rho \sigma}
\;V^a_\mu \;\partial_\nu A_\rho \; \partial_\sigma w^b \; Tr[ Q \{T^a,T^b \} ]
,
\label{eq:extra}
\end{equation}
where $\kappa$ is a constant.  This is the first time that the charge matrix
$Q$ becomes relevant. In fact, for $Q$ given by Eq.(\ref{eq:charge})
we find that the above lagrangian is proportional to $Y_L$.
It is interesting to recall that in the context of the non-relativistic
quark model one has \cite{quark} :
\begin{equation}
\frac{ A(\rho \rightarrow \pi \gamma) }{ A(\omega \rightarrow \pi \gamma) }
=  \frac{ (e_u + e_d) }{(e_u - e_d)}
\end{equation}
and there would be no $\rho \rightarrow \pi \gamma$ decay if the quark
charge matrix were given by Eq.(\ref{eq:charge}) with $Y_L = 0$.
Here we'll include the effect of the trace in an effective coupling
$\kappa'$. In QCD, $\kappa' \approx 0.03$ from radiative
vector meson decays.

With the interaction given by Eq.(\ref{eq:extra}) it is straightforward
to compute
the contribution to $\gamma \gamma \rightarrow z z$ arising from a
$t$ and $u$-channel exchange of a vector resonance  \cite{Ko}:
\[
T_{\mu \nu}^{t+u}  = i  (\frac{\kappa' e g''}{v})^2
\left\{  \left[ g_{\mu \nu} - \frac{k_{1 \nu} k_{2 \mu}}{k_1 \cdot k_2} \right]
\left( \frac{st/4}{t - M_V^2} + \frac{su/4}{u - M_V^2} \right) -  \right.
\]
\begin{equation}
\left. \left[ \frac{s}{2} p_{1 \mu} p_{2 \nu} + \frac{ut}{4} g_{\mu \nu} +
\frac{t}{2} k_{2 \mu} p_{1 \nu} + \frac{u}{2} k_{1 \nu} p_{1 \mu} \right]
\left( \frac{1}{t - M_V^2} + \frac{1}{u - M_V^2}  \right)  \right\}
\end{equation}
and finally the total amplitude in this model is given by :
\begin{equation}
T_{\mu \nu}^{V} = -\frac{\alpha}{2 \pi}
\left( g_{\mu \nu} - \frac{k_{1 \nu} k_{2 \mu} }{k_1 \cdot k_2} \right)
A_V(w^+ w^- \rightarrow z z) +
T_{\mu \nu}^{t+u} .
\end{equation}

\section{Cross section and comparison between models }
\hspace*{\parindent}

For amplitudes written as Eq.( \ref{eq:amplitude}) the cross section is given
by :
\begin{equation}
\frac{d \hat{\sigma}}{d t} = \frac{1}{128 \pi s^2} | T_{\mu \nu} |^2
\end{equation}

For the no-resonance model and for the scalar model we get :
\begin{eqnarray}
\hat{\sigma}_{NR} (s) &=& \frac{\alpha^2}{256 \pi^3} \frac{s}{v^4}   \\
\hat{\sigma}_H (s) &=& \frac{\alpha^2}{256 \pi^3} \frac{s}{v^4}
|1 - \frac{g_H^2 s}{s - M_H^2 + i M_H \Gamma_H} |^2  .
\end{eqnarray}

For the vector model we find :
\begin{equation}
\hat{\sigma}_V (s) = \frac{1}{128 \pi s^2} \; \int_{-s}^{0} \; dt \;
\left[ 2 A^2 + \frac{(u t)^2}{8} B^2  \right]   ,
\end{equation}
where
\begin{eqnarray}
A &=& -\frac{\alpha}{2 \pi} \left[ \frac{s}{4 v^2} (4 - 3 a) + \frac{a M_V^2}{
4 v^2} \left( \frac{u-s}{t-M_V^2} + \frac{t-s}{u-M_V^2} \right) \right]
\nonumber  \\
& & + (\frac{k' e g''}{v})^2 \left[ \frac{s t/4}{t-M_V^2} +
\frac{s u/4}{u - M_V^2} \right]   \\
B &=&   (\frac{k' e g''}{v})^2  \left[ \frac{1}{t-M_V^2} +
\frac{1}{u - M_V^2} \right] \nonumber
\end{eqnarray}

In Figs. $1a,1b$ we plot these cross sections for typical values of the
parameters. We immediately see that the model with a vector resonance
is characterized by an {\it absence } of the signal.
This can be understood by the fact that, in contrast to the scalar resonance
case, there is no $s$-channel
resonance contribution to the rescattering process $w^+ w^- \rightarrow
z z$ in the vector resonance model.

A few words are now
in order with respect to our calculation in the vector model.
At the energies we are considering, for $\kappa'$ close to its QCD value,
the contribution from Eq. (\ref{eq:extra}) is very small, the end result
being almost identical to taking $\kappa' = 0$. There is a priori no
dynamical reason for not having larger values of $\kappa'$ ($\kappa'$ is
proportional to the techniquark magnetic moment), which typically
results in an enhancement of the cross section at high energies.

Also of interest is the result of our calculation to QCD. In that case, we find
a cross section for $\gamma \gamma \rightarrow \pi^0 \pi^0 $ that presents
the features of interference between the two
contributions and is considerably below the experimental values. However,
when the two contributions are taken separately either
the unitarized amplitude ($ \kappa' = 0$) or the non-unitarized+vector
exchange amplitude ($a = 0$) would
fairly reproduce the QCD data. We don't expect this crude model to describe
the QCD data accurately since we have not include the effects of other
resonances like the $\omega$ and $A1$ \cite{DH} .

 We also emphasize that Eq.(\ref{eq:result})
is an {\it assumption} that, although found correct in the scalar model, may
not
be valid in the vector model. It actually introduces spurious $t-$ and
$u-$channel poles in addition to the physical ones given by the second term
in Eq.(34).
In fact, it was recently shown by Pennington and Morgan \cite{PM} that only
for models in which the amplitude for $\pi^+ \pi^- \rightarrow \pi^0 \pi^0$
is independent of the initial pions momenta does Eq.(\ref{eq:result}) holds.
This is indeed the case for the non-resonance and scalar models. They have
also shown by means of an explicit calculation that, in a model that includes
both scalar and vector resonances, Eq.(\ref{eq:result}) does not reproduce the
correct threshold behavior for $\gamma \gamma \rightarrow \pi^0 \pi^0$.
However, the {\it numerical} difference that exists between
the ansatz of  Eq.(\ref{eq:result}) and the exact result in different kinematic
regimes is still unknown.
For instance,
Dobado and Pel\'{a}ez \cite{Dob} used a Pad\'{e} unitarized amplitude for
 $\pi^+ \pi^- \rightarrow \pi^0 \pi^0$, which effectively generates a
$\rho$-meson, to compute  $\gamma \gamma \rightarrow \pi^0 \pi^0$ via
Eq.(\ref{eq:result}) and found a good agreement with their exact unitarized
result (and with data\addtocounter{footnote}{-1}
 \footnote{In their model there is no direct $\rho \gamma
\pi$ vertex, which can upset the good agreement between their results and
experimental data. }).

 A more complete calculation would involve
computing the contribution of the vector resonances in loops, that is, a
complete one-loop analysis of this process in the context of the hidden
symmetry model \cite{loop}.
Since this model is supposed to be a good description of QCD,
we don't expect any major differences from this more thorough approach.

\section{Implications for photon-photon colliders}
\hspace*{\parindent}

There has been recent interest in the possibility of producing a high energy
photon-photon collider by back-scattering a high intensity laser beam off
a high energy electron beam. Most of the scattered photons have their direction
close to the original electron beam and carry a large fraction of the initial
electron energy. The probability of finding a photon carrying an energy
fraction $x$ of the initial electron is given by \cite{laserback} :
\begin{equation}
F_{\gamma/e} (x) = \frac{N(x,\xi)}{D(\xi)}
\end{equation}
where $\xi$ is defined in terms of the electron mass $m_{e}$, the electron
initial energy $E$ and the energy of the photon in the laser beam $\omega_0$ :
\begin{equation}
\xi = \frac{4 E \omega_0}{m_e^2}
\end{equation}
and
\begin{eqnarray}
 N(x,\xi) &=& 1 - x + \frac{1}{1-x} - \frac{4 x}{\xi (1-x)} +
                                       \frac{4 x^2}{\xi^2 (1-x)^2 }  \\
D(\xi) &=& \int_0^{x_m} \; dx \; N(x,\xi)   \nonumber \\
&=& \left[ 1 - \frac{4}{\xi} - \frac{8}{\xi^2} \right] \ln (1+\xi) +
\left(\frac{1}{2} + \frac{8}{\xi} - \frac{1}{2 (1+\xi)^2}  \right)
\end{eqnarray}
where $x_m = \frac{\xi}{1+\xi}$ is the maximum energy fraction
carried by the scattered photon.

The differential cross section with respect to the invariant mass
$M_{ZZ}$ of the
final state $Z_L Z_L$ pair can be written as :
\begin{equation}
\frac{d \sigma}{d M_{ZZ}} (s) = \frac{2 M_{ZZ}}{s} \;
\frac{d{\cal L}_{\gamma \gamma}}{d \tau} (\tau) \hat{\sigma} (\tau s)
\end{equation}
where the differential photon luminosity function
$\frac{d {\cal L}_{\gamma \gamma}}{d \tau}$ is given by :
\begin{equation}
\frac{{d \cal L}_{\gamma \gamma}}{d \tau} (\tau) = \int_{\tau/x_m}^{x_m} \;
 \frac{dx}{x} \;
F_{\gamma/e} (x) F_{\gamma/e} (\tau/x) ,
\end{equation}
where $\tau = \hat{s}/s = M_{ZZ}^2/s$ .

We illustrate our results by showing in
Fig. 2  these differential cross sections for a $\gamma \gamma$ collider
originated from a $\omega_0 = 1.17 $eV laser being Compton back-scattered
from both $e^+$ and $e^-$ beams of a $1$ TeV linear collider. In this case,
$\xi = 9.4$ and $x_m = 0.90$. In this example we'll neglect $e^+ e^-$ pair
production that may occur for $x_m \geq 0.828$. We concentrate in the region
$M_{ZZ} \geq 300$ GeV, where the equivalence theorem as well as the use of
massless phase space become good approximations.
In this figure we compare the results of the vector resonance model with
different parameters with a $1$ TeV Higgs model, since for a smaller Higgs
mass the signal for the scalar model is much larger than for the vector model.
The resonant shape of the curves in this figure is an artifact arising from
the convolution of a growing point cross section with a photon luminosity
function that falls rapidly near the kinematical limit $(\sqrt{\hat{s}})_
{max} = 900$ GeV .

\section{Conclusion}
\hspace*{\parindent}

The cross section arising from the model with a vector resonance is
generally small compared to a model with a scalar resonance of mass
$M_H < 800$ GeV, where a pole structure is the dominant feature of the
signal.
However, the absence of a clear resonance structure would not rule
out the scalar model since it may be possible that the scalar mass is
above the $1$ TeV scale, in which case the cross section is small
( $\cal{O} (\mbox{fb}) $) in both models.
The situation becomes even worse when backgrounds are taken into account
\cite{Kingman}.
Recently it has been shown that the irreducible background from the
production of transversally polarized $Z$ pairs is very large
\cite{Jikia-Berger}, reducing the possibilities of finding even a scalar
resonance with mass above $300$ GeV. As techniques for the
determination of the
polarization of gauge bosons are improved and high luminosity
($ {\cal O}(100 \; \mbox{fb}^{-1} \; \mbox{year}^{-1} ) $)
$\gamma \gamma$ colliders become available \cite{vlepp} it may eventually be
possible
to extract some information on the symmetry breaking sector from the
process $\gamma \gamma \rightarrow Z_L Z_L $.
The vector resonance model could in principle be better tested at an
$e \gamma$ facility because of the $s-$channel contribution to the subprocess
$\gamma W^{+,-} \rightarrow \gamma W^{+,-}$ that exists in addition to the $W$
boson contribution. Work along these lines is in progress.

\section*{Acknowledgement}
\hspace*{\parindent}

I would like to thank H. Goldberg for
helpful discussions, and for a critical reading of the manuscript.
This research was supported in part
by the National Science Foundation under Grant No. PHY-9001439, and by the
Texas National Research Laboratory Commission under Award No. RGFY9214.

\newpage
\section*{Figure Captions}
\hspace*{\parindent}

Figure 1a : Point cross section in fentobarns as a function of the $\gamma
\gamma$ center-of-mass energy . Solid line: non-resonance model;
dotted line: $M_H = 1000$ GeV; dashed line: $M_H = 800$ GeV;
dot-dashed line: $M_H = 600$ GeV.

\bigskip

Figure 1b : Point cross section in fentobarns as a function of the $\gamma
\gamma$ center-of-mass energy . Solid line: non-resonance model;
dotted line: $M_V = 2000$ GeV, $\Gamma_V = 700$ GeV, $\kappa' = 0.2$ (cross
section is this case is scaled down by a factor of $10$);
dashed line: $M_V = 2000$ GeV, $\Gamma_V = 700$ GeV, $\kappa' = 0.03$;
dot-dashed line: $M_V = 1000$ GeV, $\Gamma_V = 300$ GeV, $\kappa' = 0.03$ .

\bigskip

Figure 2 : Differential cross sections in fb/GeV as a function of the
$Z_L Z_L$ invariant mass. Solid line:  $M_V = 2000$ GeV, $\Gamma_V = 700$ GeV,
$\kappa' = 0.03$ ; dotted line:  $M_V = 1000$ GeV, $\Gamma_V = 300$ GeV,
$\kappa' = 0.03$; dashed line:  $M_H = 1000$ GeV; dot-dashed line:
 $M_V = 2000$ GeV, $\Gamma_V = 700$ GeV, $\kappa' = 0.2$ .

\end{document}